\documentclass[conference]{IEEEtran}
\usepackage{graphicx}
\usepackage{latexsym}
\usepackage{amsmath}
\usepackage{url}
\begin{document}

\title{Toward User-Centric Feature Composition\\
for the Internet of Things}

\author{
\IEEEauthorblockN{Pamela Zave}
\IEEEauthorblockA{AT\&T Laboratories---Research\\
                  Bedminster, New Jersey, USA\\
                  pamela@research.att.com}
\and
\IEEEauthorblockN{Eric Cheung}
\IEEEauthorblockA{New York, New York, USA\\
                  cheungeric@yahoo.com}
\and
\IEEEauthorblockN{Svetlana Yarosh}
\IEEEauthorblockA{University of Minnesota\\
                  Minneapolis, Minnesota, USA\\
                  lana@umn.edu}
}
\maketitle

\begin{abstract}
Many user studies of home automation, as the most familiar representative
of the Internet of Things, have shown the difficulty of developing
technology that users understand and like.
It helps to state requirements as largely-independent features,
but features are not truly independent, so this incurs the cost of
managing and explaining feature interactions.
We propose to compose features at runtime, resolving their interactions
by means of priority.
Although the basic idea is simple, its details must be designed
to make users comfortable by balancing manual and automatic control.
On the technical side, its details must be designed to allow meaningful
separation of features and maximum generality.
As evidence that our composition mechanism achieves its goals, 
we present three substantive examples of home automation,
and the results of a user study to investigate comprehension of 
feature interactions.
A survey of related work shows that this proposal occupies a sensible
place in a design space whose dimensions include actuator type,
detection versus resolution strategies, and modularity.
\end{abstract}

\section{Introduction}
\label{sec:intro}

For twenty years the idea of
``ubiquitous'' or ``pervasive'' computing has captured the imaginations
of researchers and entrepreneurs
\cite{caceres}.
Recent improvements in wireless connectivity and inexpensive
network-enabled devices have created a surge of
interest in the (similar) concept of the
Internet of Things, in which widespread sensors and actuators connect
many previously unconnected things through the Internet 
\cite{iotcomix,atzori,iot}.
Many people
are thinking about what can be done with sensors and actuators in homes,
workplaces, schools, hospitals, public spaces, and transportation.
The number of commercial offerings is increasing rapidly.

Of all automatable environments, the home is undoubtedly the best-studied
so far.
The literature reports on a large number of ``smart home'' prototypes
and user studies, {\it e.g.,} 
\cite{aipperspach,microsoftUS,homecontrol,7challenges,homey,sabbathday}.

The results of these studies show that there is much work to be done before
the potential of the Internet of Things can be fully realized in homes,
let alone more challenging environments such as workplaces and hospitals.
In the cited papers, complaints and deficiencies far outnumber 
satisfied users.  
Typically, one hobbyist in a home plays with home automation, while
everyone else hates it \cite{homey}.

In this paper we focus on the problems of how
systems that interact constantly with people---shaping
the environments in which they live---should behave.
We know their behavior will be complex, because peoples' lives are complex.
They should do what people find intuitive and trustworthy.
They should be flexible and extensible (and possibly programmable
by the users themselves).
The help they give should be worth more than the time and trouble needed
to manage and configure the systems.
They should make 
people feel they are gaining, rather than losing, control.
As put by Edwards and Grinter,
``The challenge for smart home designers is to create systems that ensure
that users understand the pragmatics of sensors, interpretation, and
machine action as well as they understand the pragmatics of devices in
their homes now. 
From a technical perspective, the challenge of developers is . . .
to ensure that inference---when performed
at all---is done in a way that is predictable, intelligible, and 
recoverable'' \cite{7challenges}.

The complexity of these systems is a result of the diversity of their
environments and their requirements.
Requirements serve a diversity of purposes, including security,
convenience, energy conservation, fun, and independent living for the
elderly.
Requirements come from a diversity of stakeholders, including each
resident of the home, caretakers who live elsewhere, 
utility companies, and the community.
System functions are triggered by a variety of situations, including
scheduled events, human requests, sensed or predicted events, and
emergencies.
Some requirements are optional, and some evolve over time.

A requirement that is stated separately and mostly independently of
other requirements is called a {\it feature}.
It is a well-established practice to document most complex
consumer products and services in terms of features.
People can understand and remember individual features, if
they are cohesive, and can learn them incrementally.
New features can be added over time.
In this paper we will consider only systems in which features are
implemented separately as well as specified separately, because 
separate implementation extends the benefits of features to the
development and deployment process.

\begin{figure*}[!t]
\centering
\includegraphics[scale=0.80]{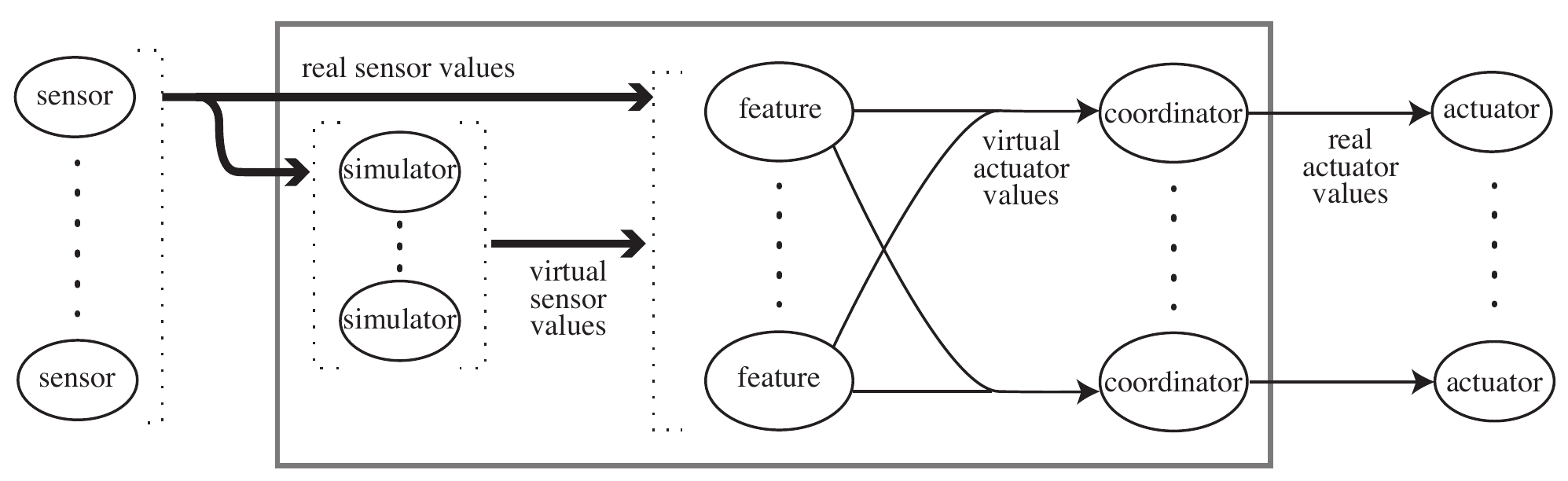}
\caption{The organization of a control system for the Internet of Things.}
\label{fig:context}
\end{figure*}

The cost of using features for specification and implementation is
that not all requirements are truly independent, and features often
interact.
A {\it feature interaction} is a conflict
or inconsistency between requirements.
Feature interactions are an implicit property of feature specifications
and the way that feature implementations
are composed to make an executable system.
Because interactions are inconsistencies, they must be resolved either by
refining the features to avoid conflicts,
or by defining a feature-composition mechanism that
automatically resolves conflicts at runtime.

This paper proposes a mechanism for runtime composition of features
that control actuators in real time.
Section~\ref{sec:context} describes the context in which this mechanism
runs, and the actuators for which it is suited.

The composition mechanism was designed with users in mind.
It is intended to have three user-centric properties:
\begin{itemize}
\item
To result in system behavior that users want.
When adding automation to human spaces, which have always been controlled
manually, it is most important to provide a good balance between
manual and fully automatic control.
\item
To be comprehensible to users.
Ultimately users must understand the feature interactions, at
least somewhat, in addition
to the features themselves.
\item
To allows simple feature specifications and implementations,
and offers guidance in avoiding mistakes.
This increases the quality and robustness of the system.
\end{itemize}
Section~\ref{sec:composition} presents the composition mechanism and 
explains the user-centric concepts in its design.

In Section~\ref{sec:evidence} we present evidence that the composition
mechanism satisfies or may satisfy these properties within its
boundaries of applicability.
The evidence is of three kinds:
\begin{itemize}
\item
Formal modeling and analysis guarantee that the mechanism has the
expected, straightforward semantics.
There is also a prototype implementation.
\item
Examples of controlling three different actuators show that
feature specifications are simple and that a wide variety of feature
functions can be handled.
\item
The results of a user study support the claims that composition results
in behavior that users want, and that feature interactions are 
comprehensible.
Equally important, the results suggest better ways to explain feature
composition to users.
\end{itemize}

As elaborated by the section on related work (Section~\ref{sec:related}),
the feature-composition
mechanism proposed in this paper represents a point in a design
space with several dimensions.
The first dimension is the kind of actuator being controlled, which
constrains what composition can do.
The second dimension is whether to emphasize 
detection or resolution (composition)
as the means of handling feature interactions.
A detection approach tends to view feature interactions as something
that must be eliminated, while a resolution approach expects them and
resolves conflicts dynamically.
A third dimension concerns modularity: whether functionality is
divided into features, and how this
affects the quality of implementations.
Section~\ref{sec:related} will show that our choices make sense from
this multi-dimensional perspective, and compare well to other work on
feature interaction.

The user-centric properties are big goals, and it will take far more
to achieve them than a single feature-composition mechanism for a
single class of actuator.
Nevertheless, it is a contribution toward reaching the goals, and
an example of how domain-specific design of a computational mechanism
can help with the challenges of user interfaces. 
Similarly, Edwards, Grinter, Mahajan, and Wetherall \cite{homenet}
stress the point that
the problems of home networking straddle human and technical
challenges, so that they cannot be solved by systems and networking
researchers focusing only on new technology, nor by HCI researchers
focusing only on the user interface.

\section{Context and examples}
\label{sec:context}

Figure~\ref{fig:context} shows the high-level organization of a
control system for the Internet of Things.
Outside the system boundary there are sensors and actuators.
Real sensor values enter the system as streams of timestamped records.
Real actuator values also leave the system as streams of timestamped
records.
We will assume for simplicity that all modules within the system
run concurrently, and communicate through streams of timestamped records,
as is standard for this kind of system
({\it e.g.,} see \cite{CQL,spitfire,perla}).

In the center of the system there are feature modules, each
implementing a separate feature as described in Section~\ref{sec:intro}.

Within the system, real sensor values are distributed to all the modules
that are interested in them.
In addition to feature modules, real sensor values are used by
{\it simulators} that 
simulate parts of the real world for the purpose of computing values
of virtual sensors.
Virtual sensors are sensors that do not exist but are needed by features.
For example, a simulator might
combine values from multiple real sensors
in a room to detect (with high probability) that there is a person in
the room.
Machine learning is often used to compute virtual sensor values.
Virtual sensor values are distributed to all the feature modules that are
interested in them, and feature modules
treat real and virtual sensor values alike.

Each feature attempts to control one or more actuators by producing a
stream of values ({\it i.e.,} commands or settings) 
for each actuator.
These are called virtual actuator values in the figure because each stream
represent the viewpoint of a single feature only.

Each actuator has a {\it coordinator} that performs feature composition
for that actuator.
It receives a merged stream of virtual actuator values from all the
features that attempt to control the actuator.
The coordinator handles the stream according to the algorithm in
Section~\ref{sec:composition}, emitting a 
stream of real actuator values that goes to the actuator.

Humans interact with the system in two ways.
They can set configuration variables to customize how features work;
these are treated as constants and not shown in Figure~\ref{fig:context}.
For real-time interaction with the system,
input devices are regarded as sensors, and output devices are
regarded as actuators.

In this paper we consider only actuators whose commands are single
values of some type.
They are called {\it settings} because when the actuator receives a
value, some part of the state of the actuator is set to that value,
overriding the previous setting.
Equally important, the actuator can 
receive and adopt a new setting at any time.
Our examples include a door lock whose settings are {\it locked}
and {\it unlocked}, a thermostat whose settings are degrees of
temperature, a furnace switch
whose settings are {\it off} and {\it on},
and a dimmer switch for lighting fixtures
whose settings are fractions between 0 and 1,
inclusive.

Although many actuators are included in this class, it excludes some
useful ones.
An output display would not be included because it can display several
messages at a time; if each actuator command contains a message to
display, a new message need not displace all previous ones.
An actuator is also excluded if a command to it initiates an action
or sequence of actions that takes time to complete.
For this kind of actuator, it would not make sense to send new commands
at arbitrary times, because the actuator might still be busy completing
another action.

Referring to Figure~\ref{fig:context},
a real sensor event (value change) can propagate through the system from
left to right, generating new virtual sensor values and fanning out to
multiple features.
Multiple features can respond by producing multiple virtual actuator
values, some of which fan in to the same coordinator.

Although it is not shown in the figure, often real actuator values on
the right are fed back to the system as sensor values on the left,
so that features can respond to how the actuators are set.
Furthermore,
the environment of the system is a simple or complex causal structure that
can feed the system's effect on actuators back to the system in the
form of new sensor values.

In the general case, this raises questions of race conditions, transient
effects, and reaching quiescence.
Although reasoning about quiescence is outside the scope of this paper,
we have done some examples showing that simple heuristics and reasoning
will suffice in many cases \cite{pfcTR}.

Concerning race conditions and transient effects, those that arise
in the environment are inevitable and the system must be programmed to
cope with them.
Inside the system, an effect (actuator value) must always be given a
later timestamp than its cause (sensor value).
If necessary, tighter synchronization may be implemented inside
the system to ensure
that modules read from input streams in timestamp order, and possibly
even that modules read all inputs with the same timestamp as a batch.
For simplicity, these implementation details will not be considered
further.

Section~\ref{sec:composition} uses the home door lock as a running
example to illustrate the concepts it introduces.
This electronic lock is fitted to the front door of a house.
There are two access panels mounted by the door, one outside the house 
and one inside.
Each panel has buttons to request that the door be locked or unlocked,
a keypad for entering passcodes, and a message display.

The requirements/features for the door lock
are typical, being drawn from real
sources.
These informal descriptions will become formal feature specifications
in Section~\ref{sec:composition}:
\begin{itemize}
\item
{\it Electronic Operation (EO):} When a person requests a lock
or unlock operation from an access panel, if that operation from that
panel requires a passcode, read and check it.
If the operation is refused, send a message to the access panel.
Otherwise, perform the requested operation. 
\item
{\it Hands-Free Entry (HFE):} When sensors detect that a resident's car
is arriving on the property, 
unlock the door so that the resident can enter
easily even carrying packages.
\item
{\it Intruder Defense (ID):} 
When sensors behind the house detect the possible
presence of an intruder, lock the door and send ``possible intruder
detected'' messages to the access panels.
Keep the door locked until the sensors provide
an ``all clear'' signal, at which time ``all clear'' messages are sent
to the access panels.
\item
{\it Night Lock (NL):} Residents configure
the time when night begins and ends.
At night, automatically lock the door and also relock it if it
becomes unlocked.
\end{itemize}
These features control the door lock actuator, and also the message
displays on both access panels.
We will consider only composition of the virtual actuator settings
for the door lock, as sequential messages to the displays do not conflict.

\section{Feature composition}
\label{sec:composition}

Like an actuator, each feature controlling an actuator has a
{\it current setting} which is its view of what the real setting should be.
When a feature's current setting changes it sends a new value to the
coordinator. 
The coordinator for an actuator
maintains a list of the current settings of all features
attempting to control that actuator.

A setting is {\it in force} if it is the coordinator's most recent
output to the actuator, and therefore the setting of the actuator.
The basic composition idea is that each feature
has a distinct numerical priority based on its importance or urgency.
At all times, the setting in force is the setting of the highest-priority
feature.

This basic idea is very simple, but it is far from complete.
It requires a number of
user-centric details and generalizations to work well.

\subsection{Some user-centric concepts}
\label{sec:users}

\vspace{2mm}
{\bf Concept:} Features correspond to situations, and should only 
constrain the actuator while the situation is occurring.

Most features are naturally
associated with particular situations that they handle.
When the situation is not occurring, the feature should have no
effect on the actuator.
For example, Intruder
Defense (ID) concerns only the situation when sensors behind the house
are detecting a possible intruder, and Night Lock (NL)
concerns only the nighttime.

To achieve this, 
it must be possible for a feature to have no current setting when
its situation is not occurring.
The type of a virtual setting must be the type
of the corresponding real setting extended with a distinguished value
{\it dontCare},
meaning that the feature does not care what the actuator setting is.
To cancel its current setting without replacing it with a new
current setting, the feature sends the pseudo-setting {\it dontCare}.
When a high-priority feature has no current setting, there is a
window of opportunity for lower-priority features to take effect.

\vspace{2mm}
{\bf Concept:} Control systems are concerned with
both manual and automatic control.

By {\it manual} control we mean actuator control that is caused by a
deliberate human action or request.
Manual control can be mechanical, such as unlocking a door with a key
or turning a light on with a mechanical switch.
Manual control can also be mediated by the computer system;
for example Electronic Operation (EO) receives user
requests through sensors,
validates them if necessary, and then responds to them electronically.

It is worth noting that manual control is usually persistent.
If a door is unlocked with a key, it stays unlocked until another
person locks it.

Automatic control occurs when the computer system infers the need to
take action on its own;
for example ID infers the possibility of an 
intruder from several different sensors, and NL locks
the door at a certain time each day.

The boundary between manual and automatic control is not always absolute.
In our running example Hands-Free
Entry (HFE) is somewhere in the middle, as the feature is triggered
by an inference that a resident is coming home.
Because the inference must be very reliable (or else the door unlocks
at unpredictable times!), and residents will come to expect it,
we classify HFE as manual control. 

The distinction between manual and automatic control is worth making,
even if it is not perfectly defined in all cases,
because the guaranteed presence of a person should affect features and
feature interaction.
This will be most clear in Section~\ref{sec:failure}.

\vspace{2mm}
{\bf Concept:}
To balance manual and automatic control, 
if all features cease to have a current setting, the real
actuator setting is left unchanged.

To satisfy its users, a control system must both allow feature composition,
and must provide a reasonable balance between manual and automatic
control.
For feature composition to work, high-priority settings must not
persist for an undetermined amount of time.
For example, the specification of EO (which will be the
highest-priority feature, see Section~\ref{sec:prio}) will say that
after a successful unlock request, the door will be unlocked for 1 minute.
This allows composition with NL, which will be the lowest-priority
feature.
During the night time, a person can unlock the door, which will stay
unlocked for 1 minute while someone passes through it.
After 1 minute the EO {\it unlocked}
setting expires, NL's low-priority setting
of {\it locked} becomes in force, and the door locks automatically.

Residents of the house can reconfigure the duration of EO if 1 minute
seems too long or short to them.
Nevertheless, it is uncomfortable for people to specify a fixed duration
for an unpredictable human situation.
If they did not have the extra features provided by home automation,
they would not need to make any such decision.

The solution to this dilemma is the rule that
if all features cease to have a current setting, the real
actuator setting is left unchanged.
This has no effect on the composition of EO and NL, because NL has a
setting of {\it locked} all night.
Consider, on the other hand, what happens during the day, when there
is no such ``default'' feature setting---and unless something happens
to trigger EO, ID, or HFE, no feature has a current setting.
If the residents wake up late, hours after the official end of nighttime,
the door will still be locked, because it was locked at nighttime and
nothing else has happened.
If a resident locks the door and leaves home, after 1 minute expires
the door will remain locked (and will stay locked until a person unlocks
it through EO or HFE).
If a resident comes home after much shopping,
and is still making trips between house and car after the 
3-minute HFE duration
has expired, the door will remain unlocked because nothing else has
happened to lock it.

Other households may prefer to keep the door locked by default at all
times, not just nighttime.
This goal is easily accomplished with a lowest-priority feature whose
setting is {\it locked} at all times.
The point of ``no setting means no change'' is that feature sets
can contain a default setting or not, as the users prefer.

\subsection{Generalization to ranges}
\label{sec:ranges}

So far we have assumed that a virtual actuator value is a
member of an enumerated set.
Even so, if two features controlling an actuator both have current
settings, the lower-priority feature may have its setting in force.
This happens whenever the current settings of the two features agree.

Some real actuator settings are
drawn from a numerical range rather than an enumerated set.
For example, a dimmer switch for lighting
fixtures may be set to any integer from 0 to 100,
each number indicating a percentage of the total light available.
For these actuators, virtual actuator settings could be subranges,
preferences (``lower is better than higher''), or possibly other
constraints.

For these actuators, the coordinator's algorithm on receiving a new
virtual setting must be generalized as follows.
After updating the priority-ordered list of current settings,
the coordinator traverses the list in descending order of priority,
accumulating constraints.
Lower-priority constraints are admitted if they are consistent with all
higher-priority constraints, and ignored if they are not consistent.
After the entire list is traversed, the controller chooses a real
actuator value that is consistent with the accumulated constraints.
As with enumerated actuator values, the highest-priority virtual setting
and possibly some lower-priority settings will be in force.
See Section~\ref{sec:examples} for features and constraints to
control a lighting dimmer switch.

\subsection{Ineffectual settings}
\label{sec:failure}

At any time, a feature's current setting is {\it ineffectual} if it is
not in force.

Although our goal is to specify and implement features as independently
as possible,
some accommodation to feature interaction is inevitable.
In our proposal feature modules learn about feature interactions
only by learning whether their
current setting is in force or ineffectual.
This is implemented by feeding replicated streams of
real actuator values from the
coordinator back to the features as if they were sensor values, as
described in Section~\ref{sec:context}.
Simpler features do not care whether their settings are in force or
not, and these can ignore the replicated streams.

\vspace{2mm}
{\bf Another user-centric concept:}
Features that implement manual control should cancel their settings
when they are immediately ineffectual or become ineffectual.

If a user requests a setting, it is acceptable for the system
to refuse the request (hopefully providing an explanation).
For example, if a user tries to unlock the door from outside but gives
the wrong passcode, EO will send a message ``invalid passcode'' to the
access panel, and not unlock the door.

What is not acceptable is for the system to ignore a manual request,
allow the user to walk away disappointed, and then honor the request
at a later time when the user may not even be aware of it.
This is what could happen if a feature implementing manual control
allows an ineffectual setting to persist.

For example, a resident of a household may drive home, triggering
HFE to unlock the door.
Around the same time, motion detectors in the back of the house
could trigger ID to lock the door.
ID has higher priority than HFE (see
Section~\ref{sec:prio}), so the door remains locked and 
HFE's {\it unlocked} setting (which will persist for 3 minutes) 
is ineffectual.

The homecoming user finds that the door is locked and sees an intruder
alert on the door panel, so goes to the back of the house, where a
squirrel has been sensed as an intruder.
Almost immediately the intrusion detection is cleared, and ID
cancels its {\it locked} setting.
The user sees that all is well and goes in the back door,
assuming that the front door is still locked.
However, 
the HFE {\it unlocked} setting now goes into force,
and the front door unlocks.
This problem would be prevented if HFE, learning that its {\it unlocked}
setting is ineffectual, cancels it.

\vspace{2mm}
In other cases, features make alternative plans on learning that
their settings are ineffectual.
For example, if the actuator can be viewed as a resource, and
there are alternative resources, the feature can try another
resource instead.

Features might also use the knowledge of when their settings are
in force or ineffectual without actually changing settings.
For example, features controlling the lighting dimmer switch might
include a feature to keep the lights on at a certain level for a
certain amount of time each day, to ensure that some houseplants
get enough light.
The houseplant feature can keep track of its minutes in force, and
cancel its setting only when the lights have been on
long enough.

\subsection{Example of feature specification}

\begin{figure*}
\centering
\includegraphics[scale=0.80]{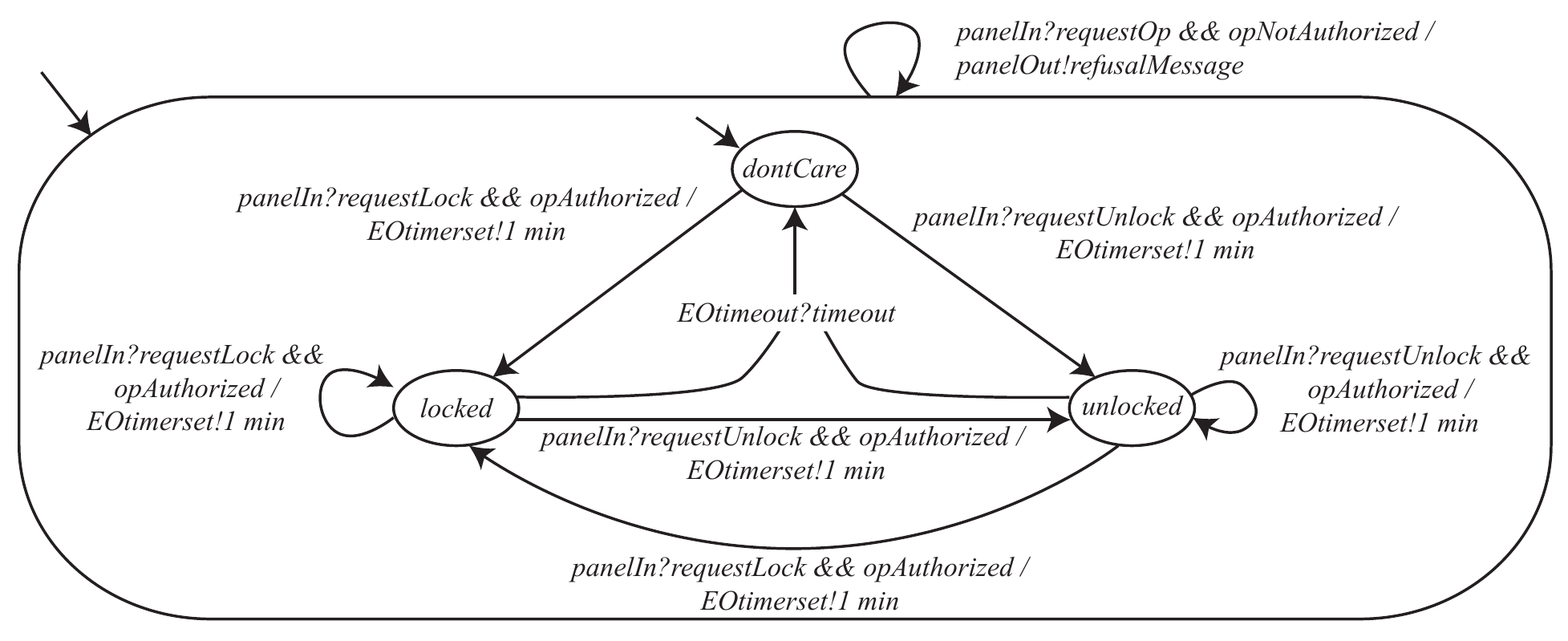}
\caption{Program for the Electronic Operation feature.}
\label{fig:EO}
\end{figure*}

A feature controlling a primary actuator
can be partially or wholly specified as a finite-state
machine whose state is the feature's current setting of the actuator
(even if feature also controls secondary actuators).
Whenever the machine changes state, the state change
implicitly sends a new setting to the
coordinator for the primary actuator.
For example, Figure~\ref{fig:EO} shows a possible specification
for Electronic Operation.
This specification can also be considered a program in a
domain-specific language.

In Figure~\ref{fig:EO}, 
reading a record from a stream is {\it streamName ? recordType},
while writing is {\it streamName ! record}.
The guard of a transition is separated from its actions by a slash.
In the figure, {\it panelOut} is a shorthand for the display of the same
panel that the corresponding request came in on.
Setting a timer and getting a timeout look like writing and reading
a stream, respectively.

Whenever there is an authorized request for a setting different from
the current setting,
the program changes state and (implicitly) sends the new setting to the
coordinator.
It also sets a local timer for 1 minute.
If there is an authorized request for the current setting, nothing is
sent to the coordinator because the program state does not change, 
but the timer is reset to a full 1 minute.
If there is a timeout, the program reverts to the
{\it dontCare} setting.

This program benefits from the simplifying assumption that it will run
at the highest priority, which means that its settings will always be
in force.
Because it implements manual control, a more robust program---one that
can run at any priority---might
check for ineffectual settings and take appropriate action.
See Section~\ref{sec:future} on future work
for further discussion of this issue.

\subsection{Priority}
\label{sec:prio}

The priority of a feature refers to its priority in controlling a
particular actuator; it may have different priorities for different
actuators.
Feature priority is encoded in signed integer values (unsigned integers
would also work).
A larger integer encodes a higher priority than a smaller integer.
In this section we mention some heuristics for assigning priorities to
features, and use them to assign priorities to the four door lock
features.

If the settings of two features can never conflict, then they do not
interact and their relative priority is irrelevant.
Two features could be non-conflicting because they have current
settings only at disjoint times.
Two features could also be non-conflicting because they never contradict
one another.  
For instance, ID and NL cannot contradict one another because they
only lock the door, never unlock it.

In general we would give features that assist manual control priority
over purely automatic control, because these are
the features for which people are waiting.
This heuristic favors EO and HFE, with EO having priority over HFE because
EO is ``more manual'' than HFE is (see Section~\ref{sec:users}).

As explained in Section~\ref{sec:users}, NL is meant to function as a
default.
Typically a default feature would have the lowest priority.

Security is important and ID should have high priority, but there are
important reasons to put its priority below that of EO.
Intrusion detection is likely to have low fidelity,
with many false positives.
Having a false positive intrusion dectection
prevent manual operation of the lock
would be very annoying.
Much more importantly, even in the case of a true positive, people
know more about the overall situation than the feature can.
If there is an intruder and he 
is chasing a resident, then people must be able to unlock
the door to let the resident in.
The resulting priority order is EO $>$ ID $>$ HFE $>$ NL.

Assigning priorities to features may not always be this easy, but
difficulty in assigning priority usually
reflects genuine complexity in the
requirements that must be dealt with one way or another.
Nevertheless, our examples show the need for another generalization.

\vspace{2mm}
{\bf Generalization to a many-to-many feature/priority association:}

So far we have assumed for simplicity that the mapping between features
controlling a particular actuator and 
priorities is one-to-one.
This is normal, but there are exceptional cases in which multiple features
must have the same priority, or a feature must send settings of the
same actuator with multiple priorities.

To implement this generalization, each virtual actuator setting is
sent to the coordinator with a {\it (feature, priority)} pair, and
the coordinator maintains a current setting for each such pair.

If there are two distinct settings at the highest priority, the 
coordinator chooses the latest one (with the most recent timestamp)
to go into force.
This means that two features using the same priority have almost the
same semantics as if they were unified into one feature in which a
more recent setting over-writes an older setting.
This is a good rule because software-development and modularity
concerns may make it necessary to merge two features or split a
feature into two.
See Section~\ref{sec:examples} for an example of two features at the
same priority.

In the same vein,
modularity
constraints might cause a feature to implement
several different requirements.
When analyzed according to heuristics, the different requirements might
need different priorities to fit into the priority order.
In these cases the different requirements can simply be implemented as 
sub-features with different priorities under one feature name.
Each sub-feature is uniquely identified by a {\it (feature, priority)}
pair.

\subsection{The composition algorithm}
\label{sec:compos}

The composition algorithm is the program run by
every coordinator.
We have seen in Section~\ref{sec:context} that records sent to a 
coordinator
must contain a timestamp {\it time} of type {\it Time}
and a {\it setting} of type {\it Setting}.
We have seen in Section~\ref{sec:prio} 
that records sent to a coordinator
must also contain a {\it feature} of type {\it Feature}
and a {\it priority} of type {\it Integer}.
These are the four fields in each record sent to a coordinator.
Except for different {\it Setting} and {\it Feature} types appropriate
to their actuators, all coordinators are alike.

The coordinator has two pieces of local state:
\begin{itemize}
\item
A list of records from the input stream, initialized
to the empty list.
This record list
contains the current settings of all unique {\it (feature, priority)}
pairs, with {\it dontCare} settings excluded.
The list is in descending priority order.
Among records with the same priority, the list is in 
descending timestamp order.
\item
A variable {\it oldSet: Setting}, whose value is the
last setting sent to the actuator.
This excludes its initial value 
{\it dontCare}, which is not sent to the actuator.
\end{itemize}
Each input record is processed in three steps, as follows.

\vspace{2mm}
{\bf Step 1: Insert record in list.}
The input record {\it matches} a record in the list if the list record
has the same feature and priority.
The cases for Step 1 are:
\begin{itemize}
\item
No matching record, input setting is {\it dontCare}: discard input record.
\item
No matching record, input setting is not {\it dontCare}: 
insert record in list at correct place for its priority and time.
\item
Matching record, input setting is {\it dontCare}: delete matching record.
\item
Matching record, input setting is not {\it dontCare}: delete matching
record, insert input record in list at correct place for its priority
and time.
\end{itemize}

\vspace{2mm}
{\bf Step 2: Choose actuator setting.}  The chosen setting is stored in
the temporary variable {\it newSet: Setting}. 
If the record list is empty, {\it newSet} $=$ {\it dontCare}.

If the list is not empty,
for a setting from an enumerated set,
{\it newSet} is the setting of its first record.
Note that the list can have more than one record with the top priority
and timestamp, in which case the choice between them based on
list order is a nondeterministic choice of setting.

If the list is not empty, for a setting from a numerical range,
the elements of {\it Setting} may be constraints or preferences
rather than values.
For constraints,
initiate a constraint set to the constraint in the first
record.
Traverse the list, adding to the set each new constraint that is 
consistent with all the previous constraints.  
The value of {\it newSet} is a subrange that satisfies all the
constraints in the accumulated set.
If there are preferences in the form of ``highest'' and ``lowest'', 
choose as {\it newSet} the highest/lowest number in the subrange,
depending on which preference has the highest priority.

\vspace{2mm}
{\bf Step 3: Decide if output needed.}  
If {\it newSet} is a value or number, is not {\it dontCare},
and is not equal to {\it oldSet},
send it to the actuator and assign its value to {\it oldSet}.
If {\it newSet} is a subrange and {\it oldSet} is not in the
subrange, choose a value from the subrange, send it to the actuator,
and assign its value to {\it oldSet}.
Otherwise, leave {\it oldSet} unchanged and set no output.

\vspace{2mm}
It is possible to make feature implementations simpler by adding more
functionality to the coordinator.
For one example, a feature could send a setting with an expiration
time.
The coordinator would set a timer for the setting, and cancel the
setting if it is still current when the timer goes off.
For another example, a feature could send a setting with an
``immediate'' tag.
The coordinator would automatically cancel this setting if it becomes
ineffectual.

\section{Evaluation of feature composition}
\label{sec:evidence}

\subsection{Formal semantics}
\label{sec:properties}

A formal model of the composition algorithm for enumerated settings
has been written in Alloy and checked with the Alloy Analyzer
\cite{alloy-book}.
The full model can be found on the Web at
\url{http://www2.research.att.com/~pamela/pfcalg.als}.
We have checked several invariants on the coordinator's record
list, along with other simple behavioral properties.
Overall behavior is characterized by the following theorem.

\vspace{2mm}
{\bf Behavior Theorem:} \\
``A feature's current setting (if any)
at a priority {\it p} and with a timestamp {\it t}
is in force unless:
\begin{itemize}
\item
some feature has a different current setting at a priority greater than
{\it p}, or
\item
some feature has a different current setting at priority {\it p}, and with
time {\it t} or greater than {\it t}.
\end{itemize}

\vspace{2mm}
The Alloy Analyzer performs
exhaustive enumeration of model instances over bounded scopes.
The invariants, behavior theorem, and other properties
have all been checked for all model instances up to 3 features,
3 priorities, 3 settings, and 6 input records to the coordinator.
Although this would not be enough to interpret as a proof of a
significant theorem, it seems sufficient to eliminate simple
flaws and establish these very basic properties.
The coordinator and some prototype features have also been implemented
in Scala, and tested without surprises.

The behavior theorem is weak because it does not answer the crucial
question of which features have which current settings at time {\it t}.
As we shall see in Section~\ref{sec:related}, this question has been
answered by other researchers with model checking.
It would be useful to check global invariants of our control
systems, {\it e.g.,} safety conditions or
properties of multiple actuators.
Model checking both could and should be applied to this problem.

On the other hand, Section~\ref{sec:related} will show that model
checking is often used to detect feature conflicts that must be removed,
or to find bugs.
With our form of feature composition it is not necessary to remove
conflicts, because they are expected and the coordinator resolves
them.
Also, features have a natural structure that helps eliminate common bugs.
Single-actuator constraints can be guaranteed by giving them 
high-enough priority.
The point of all this is that global model checking or verification
are less essential than with other approaches to feature interaction.

\subsection{Examples}
\label{sec:examples}

{\bf More on the door lock:}
Figure~\ref{fig:EO} shows the program for one of the four door lock
features.
The others are much simpler.  
For example, NL simply transitions
to a {\it locked} setting at the beginning of
nighttime, and a {\it dontCare} setting at the end of nighttime.
This illustrates that our 
automated conflict resolution is a kind of exception
mechanism that keeps features simple because they can be specified
without exceptions.

Door lock control is complicated by the fact that the
door lock can also be operated mechanically, using a key as a credential.
(This is a necessary backup mechanism, in case of power failure.)
Because of mechanical operation, there is
a sensor to tell the control system
whether the door is locked or unlocked.

Mechanical operation gives rise to a new requirement and a new feature.
Mechanical operations are manual operations as defined 
in Section~\ref{sec:users}, and
should interact with the other features in the
same way that manual electronic operations do.
For example, if the door is unlocked mechanically at night, the
door should stay unlocked for 1 minute and then be locked by NL.
It should not stay open all night, nor should it be re-locked immediately
by NL so that whoever unlocked it has no time to pass through.

\begin{figure}
\centering
\includegraphics[scale=0.80]{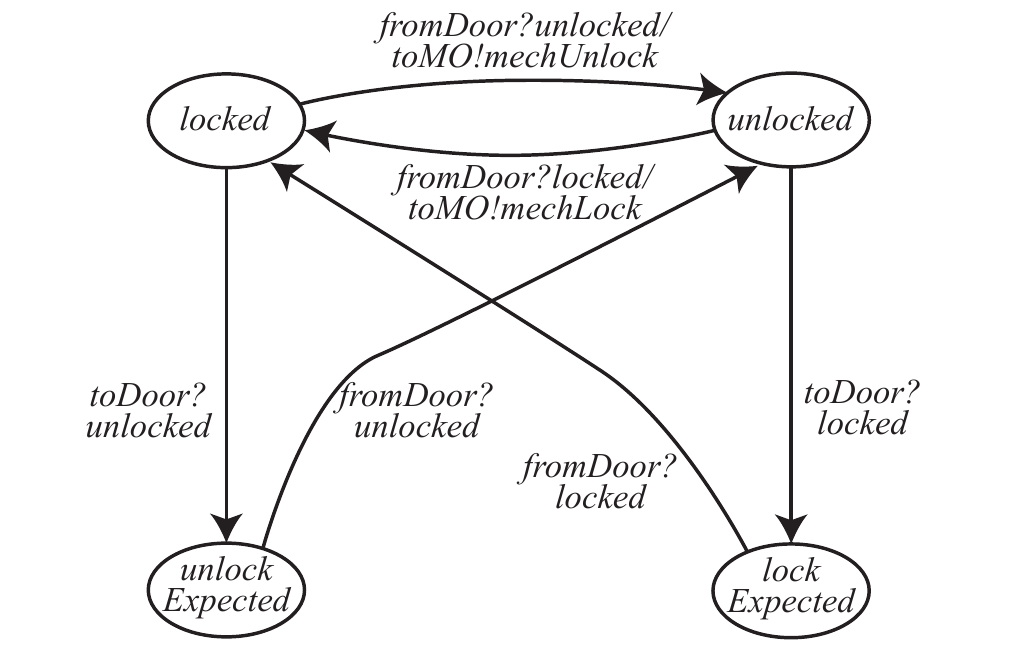}
\caption{Program for a simulator that detects mechanical lock operations.}
\label{fig:model}
\end{figure}

The first step in satisfying this requirement
is to detect mechanical operations.
This is accomplished by a {\it simulator} (as in Figure~\ref{fig:context})
that compares the actuator values sent to the door lock with the
sensor values received from it, thus detecting changes in the lock
state that are not caused by actuator values.
A program for the simulator is given in Figure~\ref{fig:model}.

The program reads two data streams:
{\it toDoor} is a replica of the real actuator values sent by the 
coordinator to the door, while
{\it fromDoor} is the stream of sensor values (encoding state changes)
from the door sensor.
For simplicity, the program assumes that lock response to the actuator
is fast and reliable,
so a new actuator value is followed by a new sensor
value within a very short time.
The program generates a stream of virtual sensor values
{\it mechLock} and {\it mechUnlock}
that is input to a new feature Mechanical Operation (MO).
Figure~\ref{fig:model} omits state initialization from sensor values.

MO has a program that is very similar to EO's.
The only difference between its program and Figure~\ref{fig:EO} is 
that a typical transition is triggered by {\it toMO ? mechUnlock}
instead of {\it panelIn ? requestUnlock \&\& opAuthorized}.
When MO receives {\it mechUnlock}, it echoes the
mechanical operation electronically by sending an {\it unlocked}
setting to the coordinator.
That record will join the record list in the coordinator,
probably causing the coordinator to produce an {\it unlocked} setting
that will be useless because the door is already unlocked.
More importantly, it will block 
lower-priority {\it locked} settings until the 1-minute duration is over.
When the 1-minute duration is over MO will send a {\it dontCare}
setting.
If this occurs at night, {\it dontCare} will stimulate the coordinator
to lock the door in accordance with the setting of NL.

Another interesting characteristic of MO is that it must have the
same priority as EO, because both forms of manual operation are
equally important, and the most recent operation should prevail.

\vspace{2mm}
{\bf Lighting dimmer switch:}
Control of
lighting a bedroom corridor in a home is described in \cite{digital-life}.
The actuator is a dimmer switch that can deliver any percentage from 0 to
100 of the total output of the light fixtures in the corridor.
The control system is sensitive to the amount of illumination coming
through a skylight in the corridor, and to motion detectors in the
corridor.
There are also several configuration variables to be set by the family.

The system has four requirements, each implemented as a separate
feature:
\begin{itemize}
\item
The {\it safety} requirement sets a minimum artificial
light level for when people
are in the corridor.
The level is different depending on whether it is the usual sleep time
or the usual awake time.
\item
The {\it health} requirement says that when residents are usually
sleeping, the
maximum light level in the corridor should be 0.  
When residents are usually
awake, there is a minimum light level in the corridor.
Because its purpose is to prevent seasonal affective disorder,
it varies for sunny and dark seasons, and also depends on how much
light is coming through the skylight.
\item
The {\it pleasantness} requirement sets a minimum level of ambient
light whenever people are likely to be in the corridor.
It depends on how much light is coming through the skylight.
\item
The {\it energy} requirement always prefers less light to more light.
\end{itemize}
These features are prioritized in the order listed (highest-priority
first).
The settings of the first three features 
are subranges of 0...100.

The feature settings are composed as in Section~\ref{sec:ranges}.
When it is the usual sleeping time and motion is detected in the
corridor, the safety requirement takes precedence and turns the lights
on.
When motion is no longer detected, the health requirement takes over
and turns the lights completely off.

\vspace{2mm}
{\bf A home furnace:}
Furnace control concerns two major actuators: a thermostat whose settings
are degrees of temperature, and an on/off switch for the furnace.

In our example, three features control the thermostat.  
{\it Manual control} enables users to set the thermostat.
{\it Learning control} is a fully automatic feature that attempts to
learn how users control the thermostat and to anticipate their actions
so that they will have less to do.
Both of these features have current settings at all times, and run
at the same priority so that each one can override the other.
At a higher priority,
{\it vacation mode} sets the thermostat to default values when the
residents are away.

In our example, four features control the furnace switch.
The {\it basic operation}
feature continually compares the house temperature with the
thermostat setting (now interpreted as a sensor value as well as an
actuator value), turning the furnace on when the house is too cold and
off when it is warm enough.
Running at higher priorities there is an {\it emergency shutoff}
feature and an {\it energy-saving} feature that makes sure the furnace
is off when the air-conditioning is on.

It is said that learning thermostats can damage heating or cooling
equipment
by cycling them on and off faster than the equipment is designed to
tolerate.
To prevent such damage, our example includes a high-priority
{\it furnace protection}
feature to guarantee that whenever the furnace is given a new setting,
no newer setting can be sent within 5 minutes (not counting an
emergency shutoff, which has higher priority).

Furnace protection is interesting for two reasons.
First, its program is conceptually very similar to that of the door
lock's MO feature.
On learning that the furnace switch has a new real setting,
It echoes that setting as its own current setting, which can only
be overridden by emergency shutoff.
After 5 minutes it cancels the setting with {\it dontCare}, so that
lower-priority features can change the real furnace setting.

Second, furnace protection is an exception to the rule, stated in
Section~\ref{sec:context}, that an actuator can adopt a new setting
at any time.
This exception, which protects the actuator or controlled object,
is conveniently specified and enforced as a feature.

\subsection{User study}

To gain some insight into user comprehension of feature interaction,
we recruited participants for a pilot user study.
The participants were all employees of a large U.S. enterprise,
with occupations in the areas of administration, law, finance,
marketing, and technology (non-research).
They were evenly distributed with respect to gender, comfort with
technology, and experience with technical troubleshooting at home.

All participants met with a researcher for an interview session.
In this session, the participant read a 350-word ``manual'' 
about the door lock
and its four features.
He or she then answered the following questions about 20 scenarios:
\begin{itemize}
\item
What is the current state of the door lock?
\item
Why do you think this?
\item
If this system were ideal, what would you {\it want} the state of the
door lock to be?
\end{itemize}
The session ended with a few more general questions.
We terminated the study after interviewing 20 participants, as we began
seeing the same patterns of responses repeat, and no new ones arise.

The results of the study were analyzed to understand the mental
models that users naturally apply to reasoning about feature
interactions.
This information may be helpful in designing control systems,
documentation or training materials, 
and further user studies.
The results of the study are reported in
detail in \cite{lanaCHI}, and we summarize some of them briefly
here.

The overall accuracy of responses was 88\% (although the answers were
binary, so random guesses would have yielded 50\%).
Gender, occupation, and comfort/experience with technology
made almost no difference in the performance of a participant, which
is encouraging because our goal is to make technology comprehensible
to everyone.

It was also encouraging that the behavior of our
feature set was in almost perfect
agreement with what most people wanted.
The only discrepancy arose from questions concerning triggering
of the Intrusion
Detection feature during a party.
Obviously it should be possible to turn the feature off during a party,
but we did not include that capability to keep the features simple
and to devise more interesting questions.

The study yielded much insight into how to explain feature composition
to users.
People understand easily that an operation such as unlocking the door
has a fixed duration such as 1 minute.
The crucial question is what happens when the minute is over:
does the lock state toggle?  return to previous?  return to a default?
not change?
Two or three simple, well-chosen examples might make the right general
answer (which depends on other features) clear to most people.

One of our final questions was, ``How would you explain the system to
a guest who is staying at your house for a week?''
The most common answer was, ``Tell them the keypad always works.''
This illustrates that simple guarantees can be achieved by
programming them into the highest-priority features.

\section{Related work}
\label{sec:related}

There is a long history of research on feature interaction and composition
in telecommunication services.
This is a more difficult problem because the controlled object, a
telephone call, has a complex actuator state.
The actions preferred by competing features take time to complete,
during which other actions must be excluded.
After an action has completed, the actuator 
state may or may not have changed
so that other competing actions are ineligible.
With these and other complications, the major solutions to the
feature-interaction problem in telecommunications are much more elaborate 
\cite{braithwaite,hayAtlee,ODFC,marplesmagill,modDFC}.

It is also possible to view home automation in terms of high-level
actions that must be completed over time \cite{homecare}.
In the remainder of this section, however, we turn to the more
typical approach to the Internet of Things, in
which actuators can change settings at any time
(as described in Section~\ref{sec:context}).

Currently many new networked
devices (``things'' in the Internet of Things) are being offered
to consumers by enterprises large and small.
As reported in \cite{tapchi}, these offers almost always promise
that users can 
program control of these
devices with ``if-then'' or ``trigger-action'' rules.
This assumption is also adopted in some research projects
\cite{mahajan,kolberg2002,tapchi}.

Trigger-action rules are independent and inherently unstructured,
so it is no surprise that even experienced programmers find
sets of trigger-action
rules difficult to write and reason about (a hobbyist says, ``. . .
it has taken me literally YEARS of these types of mistakes to iron
out all the kinks'' \cite{mahajan}).

With programs in the form of independent trigger-action rules, there
is little choice but to build tools to explore the behavior of these
programs, and hope that users will be able to provide invariants to be
checked, or detect bugs by scanning the output for 
anomalies \cite{mahajan}.

The alternative is to---in effect---group related rules into
more cohesive modules such as finite-state machines.
For example, the program in Figure~\ref{fig:EO} can be viewed as a
group of related trigger-action rules, one per state transition;
the states of the machine are values of an internal variable relating
the effects of the individual rules.
The enhanced structure provides guidelines about robust programming.
For example, a well-structured feature that turns a switch on for some
reason also includes the logic to turn it off when the reason has passed.
It is interesting to note that the self-transitions on the {\it locked}
and {\it unlocked} states in Figure~\ref{fig:EO} prevent the bug
detected in \cite{mahajan}, which is to forget to refresh a timer when
two events causing the same setting occur close together.

Some work with features or at least cohesive functional 
modules also focuses
on detecting feature interactions by finding logical conflict over
shared variables, by model checking or other means 
\cite{leelaprute,soares}.
This approach deprives users and programmers of the
feature simplification
that comes from using automatic resolution of feature interactions
as a domain-specific exception
mechanism.

The work closest to ours is 
\cite{kolberg2008},
in which 
features interact by controlling the same
object, resource, or environment variable such as air temperature.
A feature can claim exclusive access to a resource for some period
of time, some claims are not conflicting,
and priority is used to resolve conflicting claims.
It is difficult to make a closer comparison because their composition
is not completely defined, and there is no
formal semantics.
Certainly the only kind of control considered in \cite{kolberg2008}
is automatic control
as defined in Section~\ref{sec:users}, so
there is no consideration of manual control or user-centric
concepts.

Some currently available platforms for the Internet of Things include
Spitfire \cite{spitfire}, Perla \cite{perla}, and 
HomeOS \cite{homeOS,homeOSnsdi}.
Our feature composition is compatible with the goals and capabilities
of all of these, and could be implemented straightforwardly in any of
them.

\section{Future work}
\label{sec:future}

Although this approach to feature composition for the Internet of
Things looks promising, it is still preliminary, and there is much
work yet to be done.

First, there is a need to 
investigate larger examples, including unified control of 
a number of diverse actuators.
This should be done in a lab setting, with real controlled objects,
sensors, and actuators.
We already know that this approach will not work with control commands
that take time to complete.
Larger studies should hasten discovery of other limitations, showing where
additional mechanisms and integration of mechanisms are needed.

Second, we have gained just enough insight into user acceptance
and comprehension to start asking really interesting questions.
Here is a sample:
\begin{itemize}
\item
Users have simple mental models of feature interaction \cite{lanaCHI},
all of which work sometimes to explain feature composition as we
have defined it, but none of which work all the time.
Is there a way to classify and/or
re-organize situations, features, coordinators, and actuators
so that a simple mental model
suffices for each case?
\item
The door lock features take an {\it ad hoc} approach to explaining
themselves to users (sending messages to display panels), 
especially when they are interacting.
Is there an organized approach that would be better? 
\item
How do users feel about these features when they have used them for
a week, rather than talked about them for an hour?
\end{itemize}

Finally, there are questions of how static analysis,
model checking, and verification
can be used to complement runtime feature composition.
Static analysis of features might reveal the potential for
conflicts, which would feed into priority decisions as discussed
in Section~\ref{sec:prio}.
It might also be helpful in proving eventual quiescence, as discussed
in \cite{pfcTR}.
The feature structure, in conjunction with
static conflict analysis, might provide
cues that reduce the computational complexity of model checking and/or
verification for establishing global properties.
Ultimately models of the environment must participate in closed-world
reasoning about these systems.

\section{Conclusion}

While new networked devices proliferate, consumers are being sold
on a vision of the Internet of Things as easily programmed with a few
trigger-action rules.
Experience suggests that disappointment is in store for many of them.

This paper has presented an approach to specifying and implementing
complex control of the Internet of Things.
It is well-defined, as established by formal semantics.
It is quite general for controlling actuators in a particular
class, as demonstrated by numerous examples.
It allows descriptions of desired behavior to be simple and provides
guidance that reduces errors, as shown by examples and comparisons
to related work.

Most importantly, it was designed with users in mind, both in the
actuator behavior that it produces and in the way that behavior can
be explained.
A small user study indicated complete success in producing desirable
behavior, and partial success in explaining it.

Although this approach is preliminary and its results should be extended
in many directions, it is clearly 
worth pursuing.
It also illustrates the point that human and technical challenges
are so intertwined in the Internet of Things that neither can be addressed
successfully in isolation.

\section*{Acknowledgments}

This work has been improved greatly by comments from Joanne Atlee,
Josh Auzins, Greg Bond, Michael Jackson, Ratul Mahajan, and Tom Smith.

\bibliographystyle{plain}
\bibliography{pfc}

\begin{thebibliography}{10}

\bibitem{aipperspach}
Ryan Aipperspach, Ben Hooker, and Allison Woodruff.
\newblock The heterogeneous home.
\newblock In {\em Proceedings of UbiComp}, pages 222--231. ACM, 2006.

\bibitem{iotcomix}
The {I}nternet of {T}hings {C}omic {B}ook.
\newblock
  \url{http://www.alexandra.dk/uk/services/Publications/Documents/IoT_Comic_Book.pdf}.

\bibitem{CQL}
Arvind Arasu, Shivnath Babu, and Jennifer Widom.
\newblock The {CQL} {C}ontinuous {Q}uery {L}anguage: Semantic foundations and
  query execution.
\newblock {\em International Journal on Very Large Data Bases}, 15(2):121--142,
  June 2006.

\bibitem{atzori}
Luigi Atzori, Antonio Iera, and Giacomo Morabito.
\newblock The {I}nternet of {T}hings: A survey.
\newblock {\em Computer Networks}, 54(15):2787--2805, 2010.

\bibitem{braithwaite}
Kenneth~H. Braithwaite and Joanne~M. Atlee.
\newblock Towards automated detection of feature interactions.
\newblock In {\em Feature Interactions in Telecommunications Systems}, pages
  36--59, Amsterdam, 1994. IOS Press.

\bibitem{homeOS}
A.~J.~Bernheim Brush, Jaeyeon Jung, Ratul Mahajan, and James Scott.
\newblock Homelab: Shared infrastructure for home technology field studies.
\newblock In {\em Proceedings of the Workshop on Systems and Infrastructure for
  the Digital Home (HomeSys)}. ACM, 2012.

\bibitem{microsoftUS}
A.~J.~Bernheim Brush, Bongshin Lee, Ratul Mahajan, Sharad Agarwal, Stefan
  Saroiu, and Colin Dixon.
\newblock Home automation in the wild: Challenges and opportunities.
\newblock In {\em Proceedings of the ACM Conference on Human Factors in
  Computing Systems (CHI)}. ACM, 2011.

\bibitem{caceres}
Ram\'{o}n C\'{a}ceres and Adrian Friday.
\newblock Ubicomp systems at 20: Progress, opportunities, and challenges.
\newblock {\em IEEE Pervasive Computing}, 11(1):14--21, January 2012.

\bibitem{mahajan}
Jason Croft, Ratul Mahajan, Matthew Caesar, and Madan Musuvathi.
\newblock Back to the future: Forecasting program behavior in automated homes.
\newblock Technical report, Microsoft Research MSR-TR-2012-131, 2012.

\bibitem{homecontrol}
Scott Davidoff, Min~Kyung Lee, Charles Yiu, John Zimmerman, and Anind~K. Dey.
\newblock Principles of smart home control.
\newblock In {\em Proceedings of UbiComp}. ACM, 2006.

\bibitem{homeOSnsdi}
Colin Dixon, Ratul Mahajan, Sharad Agarwal, A.~J. Brush, Bongshin Lee, Stefan
  Saroiu, and Paramvir Bahl.
\newblock An operating system for the home.
\newblock In {\em Proceedings of the 9th USENIX/ACM Symposium on Networked
  Systems Design and Implementation (NSDI '12)}, 2012.

\bibitem{7challenges}
W.~Keith Edwards and Rebecca~E. Grinter.
\newblock At home with ubiquitous computing: Seven challenges.
\newblock In {\em Proceedings of UbiComp}. ACM, 2001.

\bibitem{homenet}
W.~Keith Edwards, Rebecca~E. Grinter, Ratul Mahajan, and David Wetherall.
\newblock Advancing the state of home networking.
\newblock {\em Communications of the ACM}, 54(6):62--71, June 2011.

\bibitem{iot}
Mohamed~Ali Feki, Fahim Kawsar, Mathieu Boussard, and Lieven Trappeniers.
\newblock The {I}nternet of {T}hings: The next technological revolution
  ({I}ntroduction to special issue).
\newblock {\em IEEE Computer}, 46(2):24--25, February 2013.

\bibitem{hayAtlee}
Jonathan~D. Hay and Joanne~M. Atlee.
\newblock Composing features and resolving interactions.
\newblock In {\em Proceedings of the 8th ACM SIGSOFT Symposium on Foundations
  of Software Engineering}, pages 110--119. ACM, 2000.

\bibitem{alloy-book}
Daniel Jackson.
\newblock {\em Software Abstractions: Logic, Language, and Analysis}.
\newblock MIT Press, 2006, 2012.

\bibitem{ODFC}
Michael Jackson and Pamela Zave.
\newblock Distributed {F}eature {C}omposition: A virtual architecture for
  telecommunications services.
\newblock {\em IEEE Transactions on Software Engineering}, 24(10):831--847,
  October 1998.

\bibitem{kolberg2002}
Mario Kolberg, Evan Magill, Dave Marples, and Simon Tsang.
\newblock Feature interactions in services for {I}nternet personal appliances.
\newblock In {\em Proceedings of the IEEE International Conference on
  Communications}, pages Volume 4, 2613--2618, 2002.

\bibitem{leelaprute}
Pattara Leelaprute, Takafumi Matsuo, Tatsuhiro Tsuchiya, and Tohru Kikono.
\newblock Detecting feature interactions in home appliance networks.
\newblock In {\em Proceedings of the 9th ACIS International Conference on
  Software Engineering, Artificial Intelligence, Networking, and
  Parallel/Distributed Computing}, pages 895--903. IEEE, 2008.

\bibitem{marplesmagill}
Dave Marples and Evan~H. Magill.
\newblock The use of rollback to prevent incorrect operation of features in
  {I}ntelligent {N}etwork based systems.
\newblock In {\em Feature Interactions in Telecommunications and Software
  Systems V}, pages 115--134, Amsterdam, 1998. IOS Press.

\bibitem{spitfire}
Dennis Pfisterer, Kay R{\"{o}}mer, Daniel Bimschas, Henning Hasemann, Manfred
  Hauswirth, Marcel Karnstedt, Oliver Kleine, Alexander Kr{\"{o}}ller, Myriam
  Leggieri, Richard Mietz, Max Pagel, Alexandre Passant, Ray Richardson, and
  Cuong Truong.
\newblock {SPITFIRE}: Towards a {S}emantic {W}eb of {T}hings.
\newblock {\em {IEEE} Communications}, 49(11):40--48, November 2011.

\bibitem{perla}
Fabio~A. Schreiber, Romolo Camplani, Marco Fortunato, Marco Marelli, and Guido
  Rota.
\newblock Perla: A language and middleware architecture for data management and
  integration in pervasive information systems.
\newblock {\em IEEE Transactions on Software Engineering}, 38(2):478--496,
  March 2011.

\bibitem{soares}
Christophe Soares, Rui~S. Moreira, Ricardo Morla, Jos{\'e} Torres, and Pedro
  Sobral.
\newblock Graph-based approach for interference free integration of pervasive
  applications.
\newblock In {\em Proceedings of the 7th International Symposium on Wireless
  and Pervasive Computing}, 2012.

\bibitem{homey}
Leila Takayama, Caroline Pantofaru, David Robson, Bianca Soto, and Michael
  Barry.
\newblock Making technology homey: Finding sources of satisfaction and meaning
  in home automation.
\newblock In {\em Proceedings of UbiComp}. ACM, 2012.

\bibitem{tapchi}
Blase Ur, Elyse McManus, Melwyn Pak~Yong Ho, and Michael~L. Littman.
\newblock Practical trigger-action programming in the smart home.
\newblock In {\em Proceedings of the ACM Conference on Human Factors in
  Computing Systems (CHI)}. ACM, 2014.

\bibitem{homecare}
Feng Wang and Kenneth~J. Turner.
\newblock Policy conflicts in home care systems.
\newblock In {\em Feature Interactions in Software and Communication Systems
  IX}, pages 54--65, Amsterdam, 2008. IOS Press.

\bibitem{kolberg2008}
Michael Wilson, Mario Kolberg, and Evan~H. Magill.
\newblock Considering side effects in service interactions in home
  automation---an online approach.
\newblock In {\em Feature Interactions in Telecommunications and Software
  Systems IX}, pages 172--187. IOS Press, 2008.

\bibitem{sabbathday}
Allison Woodruff, Sally Augustin, and Brooke Foucault.
\newblock Sabbath day home automation: {``}it's like mixing technology and
  religion{''}.
\newblock In {\em Proceedings of the ACM Conference on Human Factors in
  Computing Systems (CHI)}. ACM, 2007.

\bibitem{lanaCHI}
Svetlana Yarosh and Pamela Zave.
\newblock Locked or unlocked: Mental models of feature-interaction resolution
  for home automation.
\newblock Submitted for publication to CHI 2015.

\bibitem{modDFC}
Pamela Zave.
\newblock Modularity in {D}istributed {F}eature {C}omposition.
\newblock In Bashar Nuseibeh and Pamela Zave, editors, {\em Software
  Requirements and Design: The Work of Michael Jackson}, pages 267--290. Good
  Friends Publishing, 2010.

\bibitem{digital-life}
Pamela Zave.
\newblock A {D}igital {L}ife challenge problem and proposed solution.
\newblock Technical report, {AT\&T} Laboratories---Research, March 2013.

\bibitem{pfcTR}
Pamela Zave and Eric Cheung.
\newblock A modular programming abstraction for ubiquitous computing.
\newblock Technical report, {AT\&T} Laboratories---Research, March 2014.

\end{thebibliography}

\end{document}